\documentclass[12pt,a4paper,aps, prb, amsmath, amssymb]{revtex4-1} 
\usepackage{graphicx}

\usepackage{floatrow} 
\usepackage{mathptmx}      
\usepackage{amsmath}
\usepackage{setspace}
\usepackage{natbib}
\usepackage[mathcal]{euscript}

\usepackage{lineno,xcolor}

\usepackage{float}
\floatstyle{plaintop}
\restylefloat{table}

\begin{document}

\title{Vector spectroscopy for spin pumping}

\author{J. Lustikova}
 \email{lustikova@imr.tohoku.ac.jp}
 \affiliation{Institute for Materials Research, Tohoku University, Sendai 980-8577, Japan}

\author{Y. Shiomi}
 \affiliation{Institute for Materials Research, Tohoku University, Sendai 980-8577, Japan}
 \affiliation{Spin Quantum Rectification Project, ERATO, Japan Science and Technology Agency, Sendai 980-8577, Japan}

\author{E. Saitoh}
 \affiliation{Institute for Materials Research, Tohoku University, Sendai 980-8577, Japan}
 \affiliation{Spin Quantum Rectification Project, ERATO, Japan Science and Technology Agency, Sendai 980-8577, Japan}
 \affiliation{WPI Advanced Institute for Materials Research, Tohoku University, Sendai 980-8577, Japan}
 \affiliation{Advanced Science Research Center, Japan Atomic Energy Agency, Tokai 319-1195, Japan}
 
\date{\today} 
\begin{abstract}
We propose a method to separate the inverse spin Hall effect (ISHE) from galvanomagnetic effects in spin pumping experiments on metallic bilayer systems by measuring the dc electromotive force in two orthogonal directions. Calculations of dc voltages in longitudinal and Hall directions induced in Ni$_{81}$Fe$_{19}$ and  Ni$_{81}$Fe$_{19}$$\mid$Pt films at ferromagnetic resonance in a microwave cavity predict that contributions from ISHE and from the galvanomagnetic effects, i.e. the anisotropic magnetoresistance and the anomalous Hall effect, exhibit distinct signal symmetry as well as angular dependence when changing the direction of the external field with respect to the film plane. According to measurements on Ni$_{81}$Fe$_{19}$$\mid$Pt, only that dc voltage component which includes ISHE is more than five times larger than purely galvanomagnetic components. This is corroborated by results on La$_{0.67}$Sr$_{0.33}$MnO$_3$$\mid$Pt samples, demonstrating universality of this method.
\end{abstract}

\maketitle

\section{Introduction}

Knowledge of the transport properties of ferromagnetic films at ferromagnetic resonance (FMR) is vital in spin pumping experiments, where spin current (the flow of electronic spin) is injected into an adjacent paramagnetic metal upon FMR in the ferromagnetic layer.\cite{mizukami,tserkovnyak,saitoh} Spin pumping in combination with the inverse spin Hall effect (ISHE),\cite{saitoh, costache, azevedo-ishe,kajiwara,ando, ando-prb} which enables conversion of spin current into charge current, are powerful tools for exploring spin-charge related phenomena in bilayer systems.  ISHE is a relativistic phenomenon occurring in the presence of spin-orbit coupling, where both up and down electrons with opposite velocity directions are deflected into the same direction resulting in an electromotive force transverse both to the spin current flow direction $\mathbf{j}_s$ and to the spin polarization $\mathbf{\sigma}$:
\begin{equation}
\mathbf{E}_{\text{ISHE}} \propto \mathbf{j}_s \times \mathbf{\sigma}.
\end{equation}

To accurately estimate the contribution of ISHE in the observed electromotive force in systems comprising metallic ferromagnet/paramagnet bilayers, it is necessary to separate contributions from galvanomagnetic effects in the ferromagnetic layer, i.e. the anisotropic magnetoresistance (AMR)\cite{mcguire} and the anomalous Hall effect (AHE).\cite{nagaosa} The electromotive force $\textbf{E}$ induced by a current density $\textbf{J}$ flowing through a ferromagnetic metal with magnetization $\textbf{M}$ can be expressed as\cite{jan}
\begin{equation}
\textbf{E}=\rho_\perp \mathbf{J}+
\frac{\rho_\parallel - \rho_\perp}{M^2}
\left(\mathbf{M}\cdot \mathbf{J}\right) \mathbf{M}
-R_\text{a}\mathbf{J} \times \mathbf{M} \label{jan},
\end{equation}
where $\rho_{\perp}$, $\rho_{\parallel}$, $R_\text{a}$ are the resistivity in a transverse magnetic field, that in a longitudinal magnetic field, and the anomalous Hall coefficient, respectively. The first term on the right-hand side of Eq. \eqref{jan} expresses the Ohm law, the other two terms the contribution from the galvanomagnetic effects: AMR and AHE, respectively. That component of AMR which is perpendicular to $\mathbf{J}$ is often referred to as the planar Hall effect.\cite{jan}

In case of spin pumping setups,\cite{saitoh,LC,azevedo,harder} both the electric field component and the magnetic field component of the microwave can induce an rf current in the sample. Coupling of the oscillating induction current $\textbf{j}$ and of the dynamic magnetization $\textbf{m}$ at FMR excited in thin sheets of ferromagnetic conductors by a microwave gives rise to time-independent electromotive forces (the spin rectification effect).\cite{juretschke-1960,juretschke-1962} In metallic bilayers, these appear simultaneously with ISHE and obstruct the observation and estimation of pure spin injection phenomena. In recent years, significant amount of research has been done in order to devise a universal method for separating ISHE from galvanomagnetic effects in spin pumping experiments.\cite{LC, LC-apex, harder, azevedo, bai, obstbaum,weetee,haidar} The situation is complicated by the fact that the magnetic field spectrum of a voltage signal induced by AMR or AHE can have both symmetric (Lorentzian) and anti-symmetric (dispersive) contributions, and can fully overlap with the Lorentzian spectrum of the ISHE signal. \cite{azevedo,harder,LC}

In this work, we propose a method to separate ISHE from galvanomagnetic effects by investigating the angular dependence of the symmetric and anti-symmetric parts of two orthogonal components of the induced dc electromotive force at FMR in a microwave cavity while changing the direction of the magnetic field out of the film plane. Comparison of electromotive force measurements on Ni$_{81}$Fe$_{19}$ and Ni$_{81}$Fe$_{19}$$\mid$Pt samples shows that only in the bilayer samples is the component including ISHE more than five times larger than purely galvanomagnetic components, indicating dominance of ISHE.

\section{Theoretical formulation}

\begin{figure}[t]
\centerline{\includegraphics[width=8.5cm, angle=0, bb=0 0 500 500]{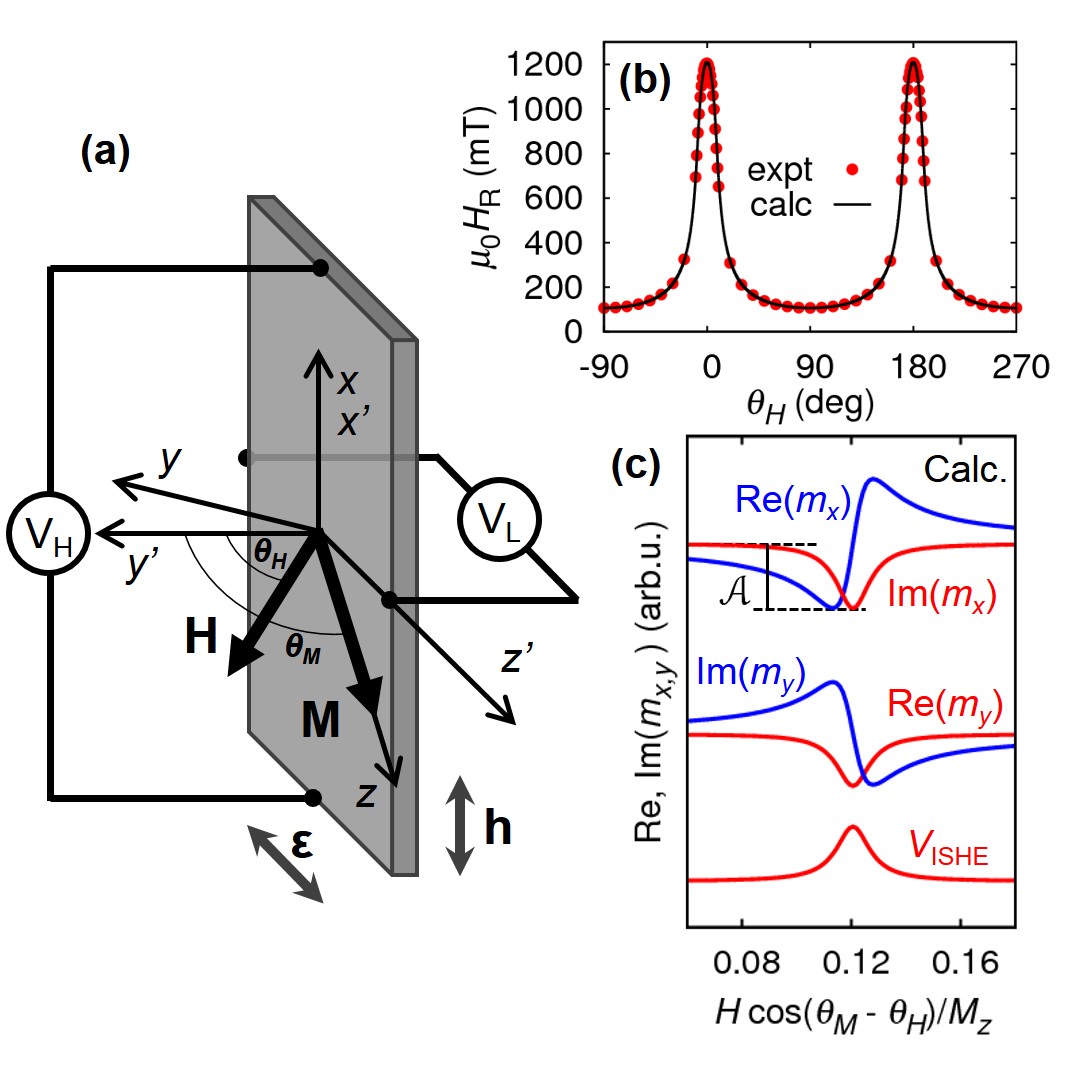}}
\caption{\label{fig1}
(a) Measurement geometry in a cylindrical TE$_{011}$ microwave cavity. Here, $\textbf{H}$ denotes the external static magnetic field, $\textbf{M}$ the static magnetization, $\textbf{h}$ the rf magnetic field, $\mathbf{\epsilon}$ the rf electric field, $\theta_M$ the angle between the surface normal and the magnetization vector, and $\theta_H$ the angle between the surface normal and the external magnetic field.  Voltage is measured along the $x'$ direction ($V_{\text{H}}$) and the $z'$ direction ($V_{\text{L}}$) as the sample is rotated about the $x$ axis.
(b) Comparison of the calculated (``calc") and measured (``expt") angular dependence of the resonance field $\mu_0 H_{\text{R}}$ with $\omega/\gamma=0.325$ T, $\mu_0 M_z = 0.884$ T, $\alpha=0.02$.
(c) Calculated spectral shapes of the real and imaginary components of the dynamic magnetization components $m_x$ and $m_y$, as well as that of the ISHE voltage [Eq. \eqref{ISHE}] with parameters as in (b). 
}
\end{figure}

We consider a metallic thin-film sample consisting of (i) a ferromagnetic single layer and (ii) a ferromagnet-paramagnet bilayer placed close to the centre of a cylindrical TE$_{011}$ microwave cavity, as illustrated in Fig. \ref{fig1}(a). The $z$ axis is taken along the magnetization vector $\textbf{M}$. Another coordinate system $x'y'z'$ is taken with the $x'$($=x$) axis along the Hall direction of the sample, and the $z'$ axis fixed along the longitudinal direction of the sample. A static external magnetic field $\textbf{H}$ is applied in the $y'z'$ plane at an angle $\theta_H$ to the surface normal $y'$. The angle between the surface normal $y'$ and the magnetization vector $\textbf{M}$ is defined as $\theta_M$.  The microwave rf magnetic field $\textbf{h}$ is oscillating along the $x$ direction,  $\textbf{h}=(h,0,0)e^{i\omega t}$, and the rf electric field $\boldsymbol{\epsilon}$ along the $z'$ direction, $\boldsymbol{\epsilon}=(0, \epsilon\cos\theta_M, -\epsilon \sin \theta_M)e^{i(\omega t + \Phi)}$.\cite{poole} Electromotive force is measured in the longitudinal ($z'$) direction and the Hall ($x$=$x'$) direction. 

The magnetization dynamics in the ferromagnetic layer is phenomenologically described by the Landau-Lifshitz-Gilbert equation,
\begin{equation}
\frac{d\textbf{M}(t)}{dt}=
-\gamma \mathbf{M}(t) \times \mu_0 \textbf{H}_{\text{eff}}+\frac{\alpha}{M}\mathbf{M}(t)\times \frac{d\textbf{M}(t)}{dt}, 
\label{LLG}
\end{equation}
where $\gamma$ is the gyromagnetic ratio and $\alpha$ the damping parameter. The time-dependent magnetization vector $\textbf{M}(t)$ is a sum of the static magnetization vector $\textbf{M}$ and the time-dependent component $\textbf{m}(t)$ oscillating in the $xy$-plane. The effective field $\textbf{H}_{\text{eff}}$ is a sum of the external field $\textbf{H}$, the demagnetizing field $\textbf{H}_M$ induced by the static magnetization vector, the rf field $\textbf{h}$ and the rf demagnetizing field $\textbf{H}_m$ induced by the dynamic magnetization.\cite{ando} Neglecting second order terms of $h$ and $m$, the solution of Eq. \eqref{LLG} is
\begin{equation}
\textbf{m}=
\left(
\begin{matrix}
m_x\\
m_y
\end{matrix}
\right)e^{i\omega t}
=\frac{1}\Xi
\left(
\begin{matrix}
\gamma^2 M_z h \left[ \mu_0H\cos(\theta_M-\theta_H)-\mu_0M_z \cos 2 \theta_M
\right]+i\alpha \gamma \omega M_z h\\
-i\gamma \omega M_z h
\end{matrix}
\right)e^{i\omega t} \label{mag},
\end{equation}
where 
\begin{align}
\Xi=&\gamma^2
\left[
\mu_0 H \cos(\theta_M-\theta_H) - \mu_0 M_z \cos 2\theta_M
\right] \nonumber\\
&\times \left[
\mu_0 H \cos(\theta_M-\theta_H) - \mu_0 M_z \cos^2 \theta_M
\right]
-(1+\alpha^2)\omega^2 \nonumber \\
&+i\alpha\omega\gamma
\left[
2\mu_0 H \cos (\theta_M-\theta_H) - \mu_0 M_z \cos 2 \theta_M -\mu_0M_z\cos^2 \theta_M
\right] \label{mag2}.
\end{align}

The ISHE electromotive force observed in the $x$ direction is proportional to the $z'$ component of the time average of the real part of $\textbf{M}(t) \times d \textbf{M}(t)/dt,$\cite{tserkovnyak, ando, ando-prb} that is,
\begin{equation}
E_{\text{ISHE}}
\propto  \overline{ \text{Re}
\left[
\textbf{M}(t) \times \frac{d \textbf{M}(t)}{dt}
\right]}| _{z'}.
\end{equation}
Using the solution of the LLG equation, one finds\cite{LC}
\begin{equation}
E_{\text{ISHE}} \propto \left[ \text{Im}(m_x)\text{Re}(m_y) -\text{Re}(m_x)\text{Im}(m_y)\right] \sin \theta_M, \label{ISHE}
\end{equation}
where the real and imaginary parts of the dynamic magnetization are defined as $\textbf{m}\equiv(\text{Re}(m_x)+i\text{Im}(m_x), \text{Re}(m_y)+i\text{Im}(m_y))e^{i\omega t}$.

The expressions for the dc electromotive forces due to AMR and AHE are obtained by finding the time average of the real part of the corresponding electric fields in Eq. \eqref{jan}:\cite{LC,mecking}
\begin{align}
\textbf{E}_{\text{AMR}} 
&\propto \overline{\text{Re} \left[ \left( \textbf{M}\cdot\boldsymbol{\epsilon} \right) \textbf{M}\right]}, \\
\textbf{E}_{\text{AHE}}
&\propto \overline{\text{Re} \left( \boldsymbol{\epsilon} \times \textbf{M} \right)}.
\end{align}
Projections into the $x$ and $z'$ axes lead to the following expressions for dc electromotive forces along the Hall direction ($E_{\text{H}}$) and the longitudinal direction ($E_{\text{L}}$), respectively:
\begin{align}
E^{\text{AMR}}_{\text{H}}&\propto \epsilon M_z \sin \theta _M \left[ \text{Re}(m_x)\cos\Phi + \text{Im}(m_x)\sin\Phi\right], \label{AMRT}\\
E^{\text{AHE}}_{\text{H}}&\propto \epsilon \sin \theta _M \left[ \text{Re}(m_y)\cos\Phi + \text{Im}(m_y)\sin\Phi\right], \label{AHET}\\
E^{\text{AMR}}_{\text{L}}&\propto \epsilon M_z \sin \theta _M \cos \theta _M \left[ \text{Re}(m_y)\cos\Phi + \text{Im}(m_y)\sin\Phi\right], \label{AMRL}\\
E^{\text{AHE}}_{\text{L}}&= 0. \label{AHEL}
\end{align}
Equations \eqref{AMRT} and \eqref{AHET} were calculated by Chen \textit{et al.} in Ref. \citenum{LC} using the same procedure. Using the approach above to calculate the contribution from the ordinary Hall effect, $\overline {\text{Re}(\mathbf{\epsilon}\times\mathbf{H}_{\text{eff}})}$, we find $E^{\text{HE}}_{\text{H}}\propto \epsilon \sin \theta _M \left[ \text{Re}(m_y)\cos\Phi + \text{Im}(m_y)\sin\Phi\right]$ 
and
$E^{\text{HE}}_{\text{L}}= 0$, which has the same angular dependence as AHE.

Treatment of the phase shift $\Phi$ requires special care. For a microwave confined in a lossless metallic resonator, the time phase shift between the rf electric field and the rf magnetic field is $\Phi=90^{\circ}$.\cite{poole} However, dissipation by currents induced in cavity walls as well as in a conducting sample and its wiring may lead to phase shifts other than $90^\circ$.\cite{azevedo} In addition, an ac current $\textbf{j}$ induced in the sample may lag behind the rf electric field $\boldsymbol{\epsilon}$ due to capacitive losses. In the calculations above, we assumed $\textbf{j} \propto \boldsymbol{\epsilon}$, where $\Phi$ is a general phase which includes the phase shift between $\mathbf{h}$ and $\mathbf{\epsilon}$, as well as that between $\mathbf{\epsilon}$ and $\mathbf{j}$.

To determine the spectral shapes and angular dependence of the electromotive forces in the case of Ni$_{81}$Fe$_{19}$ as a ferromagnetic layer, we input particular parameter values $\alpha=0.02$,\cite{ando,LC}$\omega/\gamma=0.325$ T and $\mu_0 M_z=0.884$ T. The reasonableness of the latter two parameters is demonstrated in Fig. \ref{fig1}(b) which shows the angular dependence $\theta_{H}$ of the ferromagnetic resonance field $H_{\text{R}}$. The red dots are experimental data obtained for a 10-nm-thick Ni$_{81}$Fe$_{19}$ film used in the study. The black curve was obtained by numerically solving the coupled equations for the resonance condition $(\omega/\gamma)^2=\left[ \mu_0 H_{\text{R}}\cos(\theta_H-\theta_M)-\mu_0 M_z \cos 2\theta_M\right]\times \left[ \mu_0 H_{\text{R}}\cos(\theta_H-\theta_M)-\mu_0 M_z \cos^2\theta_M\right]$ [Eq. (9) in Ref. \citenum{ando}] and for the static equilibrium condition $2 \mu_0 H \sin(\theta_H-\theta_M)+\mu_0 M_z \sin 2\theta_M=0$ [Eq. (6) in Ref. \citenum{ando}] for $H_{\text{R}}$ at each $\theta_H$. The chosen parameter values give a good agreement between calculation and experiment.

Figure \ref{fig1}(c) shows the calculated magnetic field spectra of $\text{Re}(m_x)$, $\text{Im}(m_x)$, $\text{Re}(m_y)$, and $\text{Im}(m_y)$, as well as that of $V_{\text{ISHE}}$ at $\theta_M=90^\circ$. The components $\text{Re}(m_x)$ and $\text{Im}(m_y)$ have a dispersive (anti-symmetric) spectral shape, while $\text{Im}(m_x)$ and $\text{Re}(m_y)$ exhibit an absorption-like (symmetric) spectrum.\cite{LC} As a consequence, the contributions from AHE and AMR in Eqs. \eqref{AMRT}-\eqref{AMRL} can have both symmetric and anti-symmetric components, depending on the value of the phase shift $\Phi$. By contrast, the spectral shape of the electromotive force due to ISHE is symmetric (absorption-like). In order to determine the angular dependence of each contribution we calculate the $\theta_H$ dependence of the amplitudes $\mathcal{A}$ as defined in Fig. \ref{fig1}(c).

\begin{figure*}
\centerline{\includegraphics[width=13cm, angle=0, bb= 0 0 670 450]{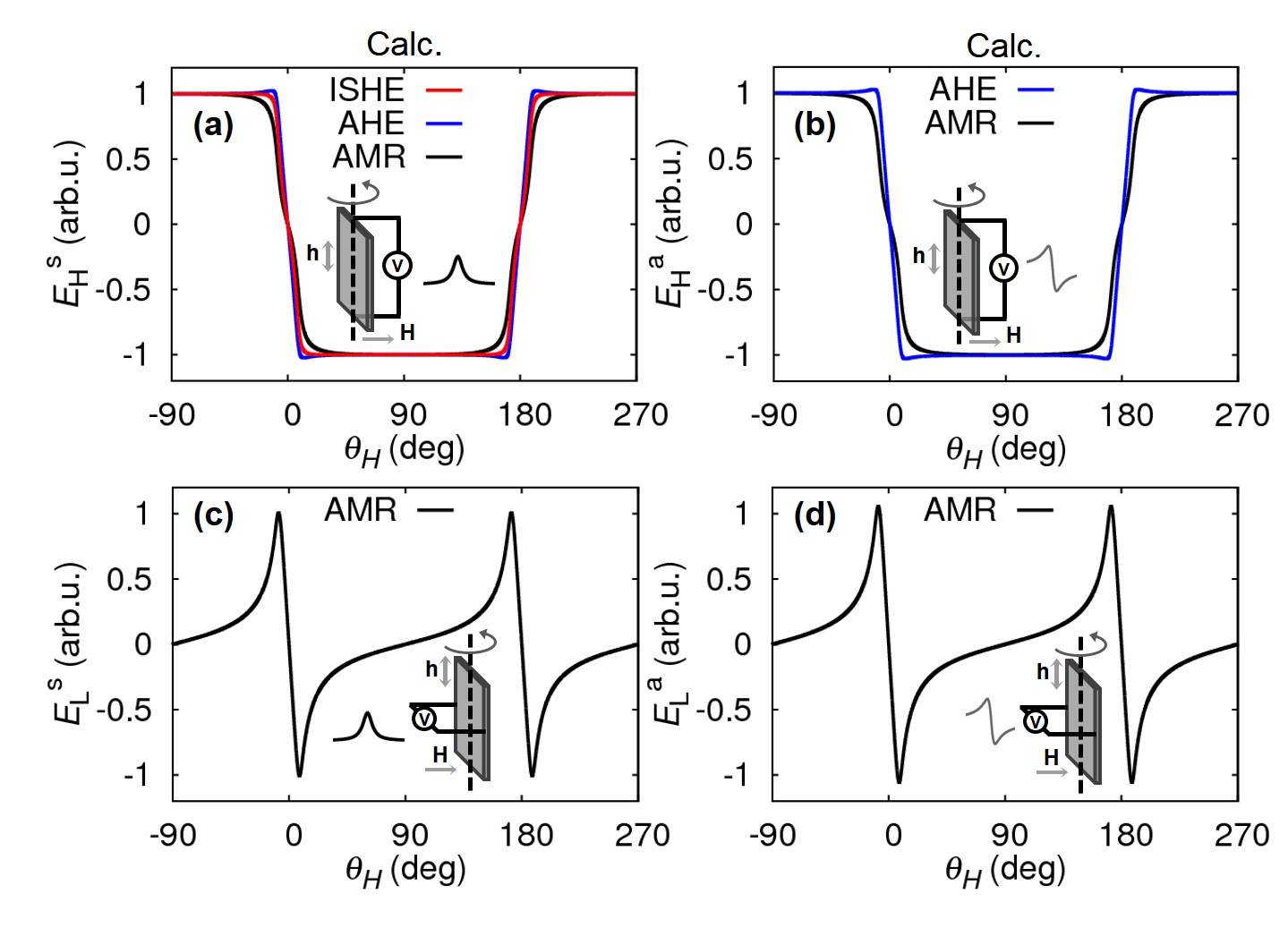}}
\caption{\label{fig2}
Calculated angular dependence of the dc-voltage amplitudes induced by the inverse spin Hall effect (ISHE), anisotropic magnetoresistance (AMR), and the anomalous Hall effect (AHE): (a) Symmetric component in the Hall direction $E_{\text{H}}^{\text{s}}$, (b) anti-symmetric component in the Hall direction $E_{\text{H}}^{\text{a}}$, (c) symmetric component in the longitudinal direction $E_{\text{L}}^{\text{s}}$, and (d) anti-symmetric component in the longitudinal direction $E_{\text{L}}^{\text{a}}$. Insets illustrate the measurement setup and signal symmetry in each case.
}
\end{figure*}

The $\theta_H$ angular dependence of the calculated amplitudes $\mathcal{A}$ of the symmetric and anti-symmetric components of ISHE, AHE and AMR in the Hall and longitudinal directions are summarized in Fig. \ref{fig2}. The contribution from ISHE [Eq. \eqref{ISHE}] appears only in the symmetric component in the Hall direction. Its angular dependence is given by $E_{\text{H}}^{\text{s}} \propto \mathcal{A}\left[ \text{Im}(m_x)\text{Re}(m_y) -\text{Re}(m_x)\text{Im}(m_y)\right] \sin \theta_M$ and is shown in Fig. \ref{fig2}(a) (red curve). It has several features: (i) The signal amplitude is maximal for in-plane magnetic field ($\theta_H=\pm90^\circ$), (ii) it is zero for perpendicular magnetic field ($\theta_H=0^\circ$, $180^\circ$), (iii) it changes sign by reversing the direction of the magnetic field, (iv) the amplitude change is abrupt around $\theta_H=0^\circ$, $180^\circ$, and (v) the signal soon saturates at its $\theta_H=90^\circ$ value when tilting the direction of the magnetic field into the film plane. 

AHE (blue curve) appears in the symmetric and anti-symmetric components of the electromotive force in the Hall direction [Eq. \eqref{AHET}]. The angular dependence of the amplitude of the symmetric component, $E_{\text{H}}^{\text{s}} \propto  \sin \theta_M \mathcal{A}\left[\text{Re}(m_y)\right] \cos \Phi$, is shown in Fig. \ref{fig2}(a) and that of the anti-symmetric component, $E_{\text{H}}^{\text{a}} \propto \sin \theta_M \mathcal{A}\left[ \text{Im}(m_y)\right] \sin \Phi$, is plotted in Fig. \ref{fig2}(b).  All of the characteristics (i)-(v) of the angular dependence of ISHE are shared by the symmetric as well as the anti-symmetric components of the AHE signal. However, here the angular dependence exhibits a cusp before saturation into the $\theta_H=90^{\circ}$ value when tilting the magnetic field from the perpendicular direction ($\theta_H=0^\circ$) into the film plane. These cusps appear at $\theta_{H}\approx \pm12^\circ$ and $180^\circ \pm12^\circ$.

Finally, contributions from AMR (black curves), are present in all four components: $E_{\text{H}}^{\text{s}}$, $E_{\text{H}}^{\text{a}}$, $E_{\text{L}}^{\text{s}}$, and $E_{\text{L}}^{\text{a}}$ [Figs. \ref{fig2}(a)-(d)]. 
The angular dependences of the amplitudes of the symmetric and anti-symmetric components in the Hall direction are given by $E_{\text{H}}^{\text{s}} \propto \sin \theta_M \mathcal{A}\left[\text{Im} (m_x)\right] \sin \Phi$ and $E_{\text{H}}^{\text{a}} \propto \sin \theta_M \mathcal{A}\left[\text{Re} (m_x)\right] \cos \Phi$, respectively [Eq. \eqref{AMRT}]. The angular dependence of $E_{\text{H}}^{\text{s}}$ [Fig. \ref{fig2}(a)] shows again all of the features (i)-(v) of ISHE. However, the change of the signal amplitude when tilting the magnetic field from the perpendicular direction to the in-plane direction is slower than for ISHE and the maximal amplitude is reached at a $\theta_H$ higher than in the case of ISHE. The same is true for $E_{\text{H}}^{\text{a}}$ [Fig. \ref{fig2}(b)]. In the longitudinal direction, AMR is the only contribution [Figs. \ref{fig2}(c) and (d)]. The angular dependences of the symmetric and anti-symmetric components are given by 
$E_{\text{L}}^{\text{s}}\propto \sin\theta_M \cos\theta_M \mathcal{A}\left[ \text{Re}(m_y)\right] \cos\Phi$ and 
$E_{\text{L}}^{\text{a}}\propto \sin\theta_M \cos\theta_M \mathcal{A}\left[ \text{Im}(m_y)\right] \sin\Phi$, respectively [Eq. \eqref{AMRL}].
Here, the angular dependences are distinctly different from those in the Hall direction with a period of $180^\circ$ and zero signal for both perpendicular and in-plane fields. The angular dependence exhibits sharp peaks of opposite polarity at $\theta_H=\pm 8.0^{\circ}$ and again at $\theta_H=180\pm 8.0^{\circ}$. This is true both for the symmetric and the anti-symmetric components and is caused by the presence of the $\sin \theta_M \cos \theta_M$ factor. We also note that due to the $\cos \Phi$ factor, the symmetric component in the longitudinal direction ($E_{\text{L}}^{\text{s}}$) is expected to be zero if $\Phi=90^\circ$.

The symmetric component of the dc voltage in the Hall direction [$E_{\text{H}}^{\text{s}}$, Fig. \ref{fig2}(a)] in general includes not only a contribution from ISHE, but also from AHE and AMR. Similarly, the anti-symmetric component in the Hall direction $E_{\text{H}}^{\text{a}}$ is a sum of the contributions from AMR and AHE [Fig. \ref{fig2}(b)]. The ratio of AMR and AHE in each case depends on the value of the phase shift $\Phi$ [Eqs. \eqref{AMRT}, \eqref{AHET}]. Therefore, knowledge of the amplitudes of $E_{\text{H}}^{\text{s}}$ and $E_{\text{H}}^{\text{a}}$ alone is not sufficient to estimate the galvanomagnetic contribution in $E_{\text{H}}^{\text{s}}$. 
 However, voltage signal in the longitudinal direction provides the missing pieces of information [Eqs. \eqref{AMRL}, \eqref{AHEL}]. The ratio of the anti-symmetric and symmetric components in the longitudinal direction, $E_{\text{L}}^{\text{a}}/E_{\text{L}}^{\text{s}}$, which are purely due to AMR, uniquely determines the phase shift $\Phi$. By contrast, the sum of the second squares,  $\left(E_{\text{L}}^{\text{a}}\right)^2+\left(E_{\text{L}}^{\text{s}}\right)^2$, determines the AMR coefficient. This information can be used to reconstruct the angular dependence of the anti-symmetric AMR signal in the Hall direction [Eq. \eqref{AMRT}], and consequently to attribute the remaining $E_{\text{H}}^{\text{a}}$ signal to AHE [Eq. \eqref{AHET}], finding the anomalous Hall coefficient. In this way, the AMR and  AHE contributions to $E_{\text{H}}^{\text{s}}$ can be in principle determined, making it possible to quantify the size of the ISHE signal.

\section{Experimental results and discussion}

Experiments were performed on Ni$_{81}$Fe$_{19}$ (10 nm) single layers and Ni$_{81}$Fe$_{19}$ (10 nm)$\mid$Pt (10 nm) bilayers. The Ni$_{81}$Fe$_{19}$ films were prepared by electron beam evaporation on a thermally oxidized Si substrate. In case of bilayer samples, Pt was deposited on the Si substrate by rf sputtering before Ni$_{81}$Fe$_{19}$ deposition. The samples were of $1\times 1.6$ mm$^2$ rectangular shapes. For spin pumping measurements, four voltage electrodes were made by wire bonding with an aluminium wire at the midpoints of the sample edges, as suggested in Fig. \ref{fig1}(a). The sample on a quartz rod was then placed close to the centre of a cylindrical cavity in TE$_{011}$ mode with a resonance frequency of $9.45$ GHz. The $Q$-values in case of a wired Ni$_{81}$Fe$_{19}$ sample and a wired Ni$_{81}$Fe$_{19}$$\mid$Pt sample were 8700 and 9500, respectively. Electromotive force measurements were performed at a microwave power $P_{\text{MW}}=200$ mW (rf field $\mu_0 h=0.16$ mT) in the Hall and longitudinal directions while rotating the sample about the Hall direction [Fig. \ref{fig1}(a)].

\begin{figure}[b]
\centerline{\includegraphics[width=6cm, angle=0, bb= 0 0 300 500]{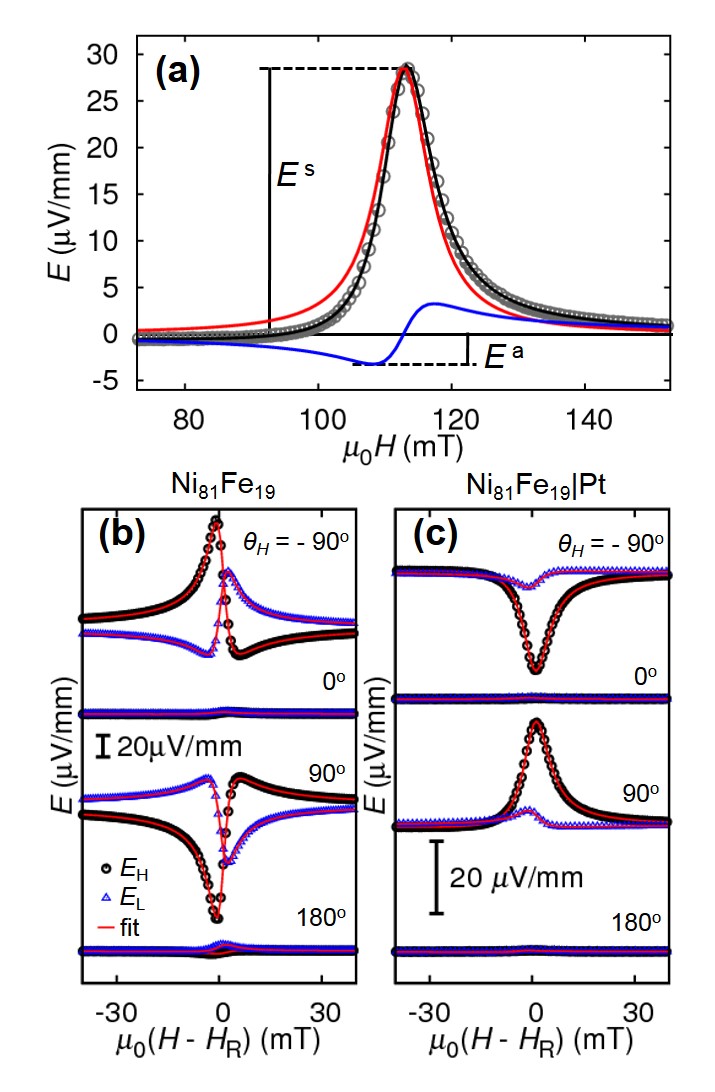}}
\caption{\label{fig3}
(a) Voltage signal in a Ni$_{81}$Fe$_{19}$$\mid$Pt sample in the Hall direction at $\theta_{H}=90^{\circ}$ (grey dots) fitted with Eq. \eqref{separation} (black curve). The red and blue curves indicate the symmetric and anti-symmetric components, respectively. $E^{\text{s}}$ and $E^{\text{a}}$ denote the amplitude of the symmetric and the anti-symmetric components.
(b), (c) Normalized voltage signals for different values of $\theta_H$ measured on a Ni$_{81}$Fe$_{19}$ (10 nm) sample[(b)], and a Ni$_{81}$Fe$_{19}$ (10 nm) $\mid$ Pt (10 nm) sample [(c)] in the Hall direction ($E_{\text{H}}$, black circles) and the longitudinal direction ($E_{\text{L}}$, violet triangles) fitted with Eq. \eqref{separation} (red curves). 
}
\end{figure}

A typical voltage signal from a Ni$_{81}$Fe$_{19}$$\mid$Pt sample measured in the Hall direction is shown in Fig. \ref{fig3}(a). The size of the voltage signal $V$ is normalized by the electrode distance $l$, $E\equiv V/l$. A voltage signal appears around the ferromagnetic resonance field when sweeping the magnetic field, containing both a symmetric (Lorentzian) component and an anti-symmetric (dispersive) component. The amplitudes of the symmetric and anti-symmetric components, $E^{\text{s}}$ and $E^{\text{a}}$, respectively, are determined by fitting the voltage signal with the following function.

\begin{equation}
E=E^{\text{s}} \frac{\Delta^2}{(H-H_{\text{R}})^2+\Delta^2}
+E^{\text{a}} \frac{-2\Delta (H-H_{\text{R}})}{(H-H_{\text{R}})^2+\Delta^2}
\label{separation}
\end{equation}

Figures \ref{fig3} (b) and (c) show the voltage signals in Ni$_{81}$Fe$_{19}$ and Ni$_{81}$Fe$_{19}$$\mid$Pt, respectively, for selected values of $\theta_H$ in both the Hall direction ($E_{\text{H}}$, black circles) as well as the longitudinal direction ($E_{\text{L}}$, violet triangles). A signal consisting of a symmetric and an anti-symmetric component is observed in all cases. The signal vanishes when the magnetic field is perpendicular to the surface plane. A sign reversal is observed when reversing the direction of the magnetic field. In all cases, good agreement is obtained by fitting the data with Eq. \eqref{separation} (red curves). The amplitude of the signal in Ni$_{81}$Fe$_{19}$ is three times larger than that in Ni$_{81}$Fe$_{19}$$\mid$Pt. This is due to shunting of galvanomagnetic effects in Ni$_{81}$Fe$_{19}$ when measuring the electromotive force in a bilayer sample, as well as due to a decrease in the microwave absorption intensity in the Ni$_{81}$Fe$_{19}$$\mid$Pt sample.

\begin{figure}[t]
\centerline{\includegraphics[width=11.5cm, angle=0, bb= 0 0 410 580]{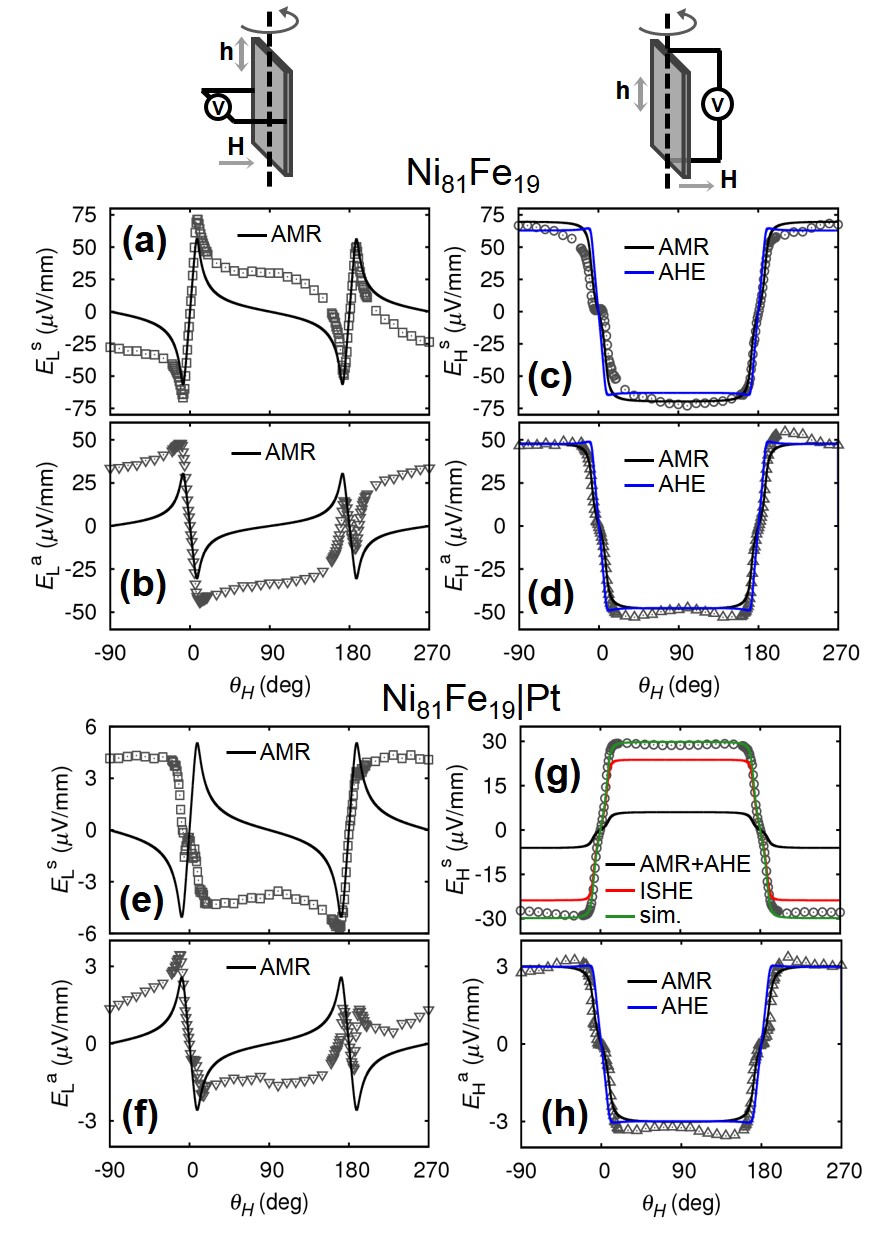}}
\caption{\label{fig4}
Measured angular dependence (grey points) of the longitudinal symmetric ($E_{\text{L}}^{\text{s}}$), the longitudinal anti-symmetric ($E_{\text{L}}^{\text{a}}$), the Hall symmetric ($E_{\text{H}}^{\text{s}}$) and the Hall anti-symmetric ($E_{\text{H}}^{\text{a}}$)  component of the electromotive force induced in a Ni$_{81}$Fe$_{19}$ sample [(a), (b), (c), and (d), respectively], and in a Ni$_{81}$Fe$_{19}$$\mid$ Pt sample [(e), (f), (g), and (h), respectively]. Black and blue curves are the calculated curves of AMR and AHE, respectively (not to scale). Red and black curves in (g) show a separation of the $E_{\text{H}}^{\text{s}}$ signal into 80\% ISHE and 20\% AMR and AHE contributions, respectively, using the calculated curves. The green curve is a sum of both contributions.
Schematic drawings at the top visualise the measurement direction. 
}
\end{figure}

The angular dependences of $E_{\text{L}}^{\text{s}}$, $E_{\text{L}}^{\text{a}}$, $E_{\text{H}}^{\text{s}}$, and $E_{\text{H}}^{\text{a}}$ extracted from fitting are shown in Figs. \ref{fig4} (a)-(h). The symmetric and anti-symmetric components in the longitudinal direction, $E_{\text{L}}^{\text{s}}$ and $E_{\text{L}}^{\text{a}}$, for a Ni$_{81}$Fe$_{19}$ sample are shown in Figs. \ref{fig4}(a) and (b), respectively. Both are plotted along with the calculated dependence (black curve, not to scale) which predicts that the only contribution comes from AMR. In case of the experimental data for $E_{\text{L}}^{\text{s}}$, the appearance of two sharp peaks of opposite polarity at $\theta_{H}=\pm8^\circ$ and again at $\theta_{H}=180\pm8^\circ$ agrees well with the prediction. However, the $\theta_H$ dependence does not fully overlap with the predicted curve. The experimental result includes a contribution which changes sign by reversing the direction of $\textbf{H}$, similar to the predicted AMR or AHE contributions to $E_{\text{H}}^{\text{s}}$. In the case of $E_{\text{L}}^{\text{a}}$ in Fig. \ref{fig4}(b), the angular dependence also includes a signal which matches the predicted AMR dependence, and a contribution from a signal which changes sign by reversing the direction of $\textbf{H}$. $E_{\text{L}}^{\text{s}}$ and $E_{\text{L}}^{\text{a}}$ have opposite polarity. 

The symmetric and anti-symmetric components in the Hall direction, $E_{\text{H}}^{\text{s}}$ and $E_{\text{H}}^{\text{a}}$ for a Ni$_{81}$Fe$_{19}$ sample are shown in Figs. \ref{fig4}(c) and (d), respectively. Both are plotted along with the predicted dependence of AMR and AHE (black and blue curves, respectively). The experimental plot for both components qualitatively agrees with the predicted dependence: the signal changes sign by reversing the direction of $\textbf{H}$ and is zero for $\theta_H=90^\circ$. Both $E_{\text{H}}^{\text{s}}$ and $E_{\text{H}}^{\text{a}}$ are of the same polarity.  In each case, there is a slight discrepancy between the experimental data and the calculated curves for AMR and/or AHE. The difference between the calculated curves for AHE and AMR is too minute to determine their ratio within the accuracy of this experiment.

The results for a Ni$_{81}$Fe$_{19}$$\mid$Pt sample are shown in Figs. \ref{fig4}(e)-(h). The symmetric component of the voltage signal in the longitudinal direction $E_{\text{L}}^{\text{s}}$ in Fig. \ref{fig4}(e) changes sign by reversing the direction of $\textbf{H}$ and has an angular dependence rather similar to that of AMR in the Hall direction. This is in disagreement with the predicted AMR behaviour in the longitudinal direction (black curve). The anti-symmetric component $E_{\text{L}}^{\text{a}}$ in Fig. \ref{fig4}(f) contains a contribution similar to the predicted contribution of AMR to $E_{\text{L}}^{\text{a}}$ (black curve) and a contribution which changes sign by reversing the direction of $\textbf{H}$. Overall, the measured angular dependence for $E_{\text{L}}^{\text{a}}$ is similar to the same component in the case of Ni$_{81}$Fe$_{19}$ [Fig. \ref{fig4}(b)]. 

 The symmetric and anti-symmetric components in the Hall direction, $E_{\text{H}}^{\text{s}}$ and $E_{\text{H}}^{\text{a}}$, for Ni$_{81}$Fe$_{19}$$\mid$Pt are shown in Figs. \ref{fig4}(g) and (h), respectively. Both change sign by reversing the direction of $\textbf{H}$ and are zero for $\theta_H=0^\circ$. Their behaviour qualitatively agrees with the predicted angle dependence for ISHE, AMR or AHE. $E_{\text{H}}^{\text{s}}$ and $E_{\text{H}}^{\text{a}}$ are of opposite polarity. In Fig. \ref{fig4}(h) the experimental data of $E_{\text{H}}^{\text{a}}$ are plotted along with the predicted angular dependence of AMR and AHE (black and blue curves, respectively). 

Two points can be noted from the observed angular dependence in the longitudinal direction. First, if $\Phi=90^\circ$, the contribution of AMR to $E_{\text{L}}^{\text{s}}$, which does not change sign by reversing the direction of $\textbf{H}$, theoretically becomes zero [Eq. \eqref{AMRL}]. The fact that a non-zero $E_{\text{L}}^{\text{s}}$ signal appears [Figs. \ref{fig4}(a) and (e)] indicates that the presence of a conducting sample and/or electrode wiring in the microwave cavity causes dissipative losses and a non-$90^\circ$ phase shift, in agreement with other reports.\cite{azevedo} Second, the presence of a signal which changes sign with reversal of $\textbf{H}$ direction in $E_{\text{L}}$ indicates that the distribution of the microwave electromagnetic fields contains components other than $h$ oscillating in the $x'$-direction and $\epsilon$ oscillating in $z'$ direction. A "contamination" of the $E_{\text{L}}$ signal from the Hall direction due to $E_{\text{H}}$ and $E_{\text{L}}$ measurement directions not being exactly orthogonal in the experimental setup does not seem to be a sufficient explanation. The reason is that the magnitude of this "contamination" signal in the longitudinal direction is comparable with that of $E_{\text{H}}^{\text{a}}$ for both Ni$_{81}$Fe$_{19}$ and Ni$_{81}$Fe$_{19}$$\mid$Pt samples.  The direction across the electrodes would need to be tilted by 45 degrees to achieve this level of mixing, while in the experiment, the tilt was a few degrees. Another aspect to consider is the tilt of the quartz rod on which the sample is fixed, which would cause a change of electric field with rotation angle. However, a simple $\propto \sin \theta_M$ or $\propto \cos \theta_M$ multiplication of $\sin \theta_M \cos \theta_M$ will not produce the observed $\sin \theta_M$-like angular dependence. Recently, magnonic charge pumping\cite{ciccarelli} has been considered as a source of charge current in ferromagnetic layers with spin-orbit coupling and magnetization dynamics. According to Ref. \citenum{azevedo2015}, spin rectification due to magnonic charge pumping is one third of the total galvanomagnetic signal in 10-nm Ni$_{81}$Fe$_{19}$ films. We have considered spin rectification due to magnonic charge current $\mathbf{j} \propto \boldsymbol{\Lambda} \cdot \partial \mathbf{m}/\partial t$ with an isotropic matrix $\boldsymbol{\Lambda}$, which parametrizes the spin-orbit coupling. AMR in this case is zero. The contribution from AHE has only a longitudinal symmetric component with an angular dependence proportional to $\sin \theta_M$. This still cannot explain a $\sin \theta_M$ dependence in the longitudinal anti-symmetric component. Thus, the most plausible explanation seems to be an altered electric field distribution due to different boundary conditions in the presence of a conducting sample and its wiring in the cavity. To remove the field distribution artefacts from the measurement, minimizing the sample volume while  maximizing the  sample-electrode surface aspect ratio might be useful. Use of high-resistivity materials in sample and electrodes might be also beneficial.

\begin{table}[tb]
\caption{Comparison of the ratios of the maxima of the electromotive force observed in the Hall ($E_{\text{H}}$) and the longitudinal ($E_{\text{L}}$) directions for the anti-symmetric ($^{\text a}$) and symmetric ($^{\text s}$) components in a Ni$_{81}$Fe$_{19}$ and a Ni$_{81}$Fe$_{19}$$\mid$Pt sample.}
\centering 
\begin{tabular}{cc cc c }
\hline
\hline                        
 & &Ni$_{81}$Fe$_{19}$& &Ni$_{81}$Fe$_{19}$$\mid$Pt
\\[0.5ex]
\hline
$E_{\text H}^{\text{a}}/E_{\text L}^{\text{a}}$ & & 1.12 & & 1.06
\\
$E_{\text H}^{\text{s}}/E_{\text L}^{\text{s}}$ & &1.02 & & 5.26
\\
 [1ex]
\hline
\end{tabular}
\label{tab1}
\end{table}

The disagreement between the experimental and calculated angular dependences of $E_{\text{L}}^{\text{s}}$ and those of $E_{\text{L}}^{\text{a}}$ obstructs the estimation of the phase shift $\Phi$ and/or the AMR coefficient which could be used to find the contribution of the galvanomagnetic effects in the Hall direction. However, there is a noticeable feature when comparing the maxima of the electromotive force in both samples. The maxima of all four components $E_{\text{L}}^{\text{s}}$, $E_{\text{L}}^{\text{a}}$, $E_{\text{H}}^{\text{s}}$, and $E_{\text{H}}^{\text{a}}$ in case of Ni$_{81}$Fe$_{19}$ are of the same order (50 to 75 $\mu$V/mm). The maximum of $E_{\text{H}}^{\text{s}}$ for Ni$_{81}$Fe$_{19}$$\mid$Pt ($30$ $\mu$V/mm) is one order of magnitude larger than the remaining three components (3 $\mu$V/mm). A comparison of the maxima of the electromotive forces in the Hall direction ($E_{\text{H}}$) and the longitudinal direction ($E_{\text{L}}$) is summarized in Tab. \ref{tab1}. In case of the anti-symmetric component in a Ni$_{81}$Fe$_{19}$ sample, the ratio of the maximum voltage signals observed in the angular dependence in the Hall direction and that in the longitudinal direction, $E_{\text{H}}^{\text{a}}/E_{\text{L}}^{\text{a}}$, is almost one ($E_{\text{H}}^{\text{a}}/E_{\text{L}}^{\text{a}}=1.12$). This is also true for the Ni$_{81}$Fe$_{19}$$\mid$Pt sample where $E_{\text{H}}^{\text{a}}/E_{\text{L}}^{\text{a}}=1.06$. The situation is different for the symmetric component. Here, $E_{\text{H}}^{\text{s}}/E_{\text{L}}^{\text{s}}=1.02$ for Ni$_{81}$Fe$_{19}$, and $5.26$ for Ni$_{81}$Fe$_{19}$$\mid$Pt. Only that $E_{\text{H}}/E_{\text{L}}$ ratio which includes a contribution from ISHE is significantly higher than the other ratios. Therefore, it is possible to attribute the increase in $E_{\text{H}}^{\text{s}}/E_{\text{L}}^{\text{s}}$ in Ni$_{81}$Fe$_{19}$$\mid$Pt to ISHE. This gives an ISHE to galvanomagnetic effects ratio of 1:0.25. The decomposition of the $E_{\text{H}}^{\text{s}}$ signal into ISHE and galvanomagnetic effects is illustrated in Fig. \ref{fig4}(g), with ISHE (red curve) constituting 80\%, and AMR and AHE (black curve) constituting 20\% of the signal at $\theta_H=90^\circ$. The sum of these two contributions is plotted as a green curve.

\begin{figure}[t]
\centerline{\includegraphics[width=16cm, angle=0, bb= 0 0 520 260]{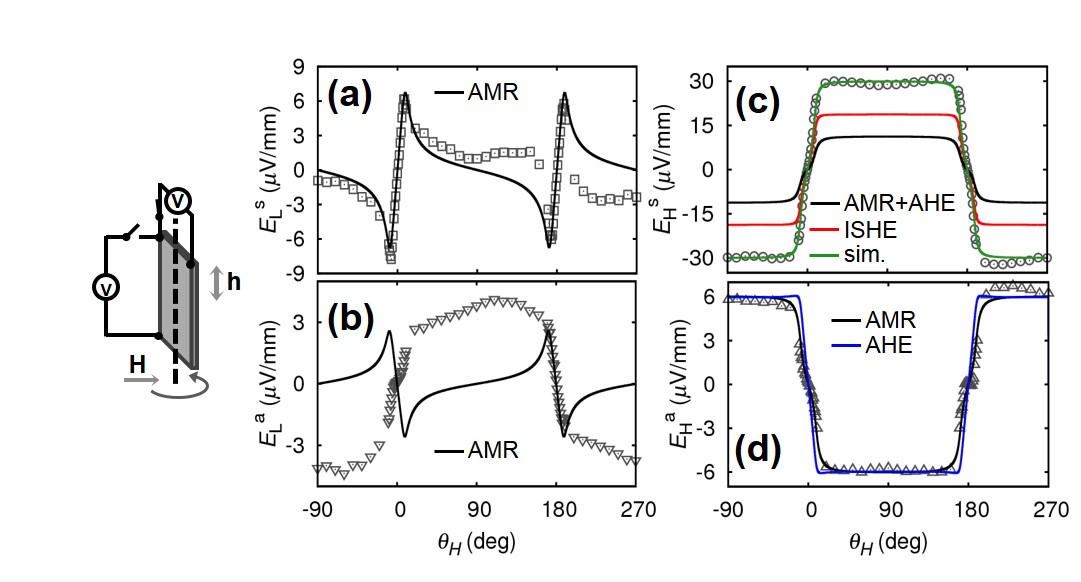}}
\caption{\label{fig5}
Measured angular dependences of the four components (a) $E_{\text{L}}^{\text{s}}$, (b) $E_{\text{L}}^{\text{a}}$, (c) $E_{\text{H}}^{\text{s}}$, and (d) $E_{\text{H}}^{\text{a}}$ for a Ni$_{81}$Fe$_{19}$$\mid$ Pt bilayer wired with three wires in the corners of the sample. The black and blue curves in (a), (b), and (d) are the calculated curves of AMR and AHE, respectively (not to scale). Red and black curves in (c) show the separation of the signal into 63\% ISHE and 37\% galvanomagnetic effects, as well as the sum of these two contributions (green curve). Scheme at the left illustrates the wiring and measurement setup.
}
\end{figure}

A different wiring does lead to slightly different $E_{\text{H}}/E_{\text{L}}$ ratios. However, the dominance of the $E_{\text{H}}^{\text{s}}$ component in Ni$_{81}$Fe$_{19}$$\mid$Pt is preserved. Figures \ref{fig5}(a)-(d) show the angular dependences measured on exactly the same Ni$_{81}$Fe$_{19}$$\mid$Pt sample as above, wired with three aluminium wires bonded in the corners of the sample, as illustrated in Fig. \ref{fig5}. The $Q$-value in this case was 9500, identical to that in the  earlier setup. 
The $E_{\text{L}}^{\text{s}}$ [Fig. \ref{fig5}(a)] and $E_{\text{L}}^{\text{a}}$ [Fig. \ref{fig5}(b)] components both contain contributions from a signal which reverses sign together with the direction of $\mathbf{H}$.  However, the polarity of $E_{\text{L}}^{\text{s}}$ and $E_{\text{L}}^{\text{a}}$ is opposite to that shown in Figs. \ref{fig4}(e) and (f), respectively. In addition, the predicted contribution from AMR in $E_{\text{L}}^{\text{s}}$ is significantly enhanced, while it almost disappears in the $E_{\text{L}}^{\text{a}}$ signal. These changes in the polarity and angular dependences in the longitudinal direction may reflect changes in the distribution of the microwave electromagnetic fields with different wiring, the former being related to the phase shift $\Phi$.
The angular dependence, magnitude, and polarity of the $E_{\text{H}}^{\text{s}}$ [Fig. \ref{fig5}(c)] and $E_{\text{H}}^{\text{a}}$ [Fig. \ref{fig5}(d)] components agree with those observed in the previous setup [Figs. \ref{fig4}(g) and (h), respectively]. 
The $E_{\text{H}}^{\text{s}}$ component visibly dominates over $E_{\text{H}}^{\text{a}}$, $E_{\text{L}}^{\text{s}}$ and $E_{\text{L}}^{\text{a}}$, the latter three being of the same magnitude. The ratios obtained were $E_{\text{H}}^{\text{a}}/E_{\text{L}}^{\text{a}}=1.54$ and $E_{\text{H}}^{\text{s}}/E_{\text{L}}^{\text{s}}=4.16$.
For comparison, the ratios for a Ni$_{81}$Fe$_{19}$ sample with this wiring were $E_{\text{H}}^{\text{a}}/E_{\text{L}}^{\text{a}}=2.28$ and $E_{\text{H}}^{\text{s}}/E_{\text{L}}^{\text{s}}=0.36$. We obtain an ISHE contribution of 63\% if comparing only $E_{\text{H}}^{\text{s}}/E_{\text{L}}^{\text{s}}$ and $E_{\text{H}}^{\text{a}}/E_{\text{L}}^{\text{a}}$ ratios for Ni$_{81}$Fe$_{19}$$\mid$Pt, which is still dominant over galvanomagnetic effects but slightly lower than in the previous setup. The increased contribution from galvanomagnetic effects can be explained by increased microwave electric field intensity caused by different wiring and/or measurement error. A breakdown of the $E_{\text{H}}^{\text{s}}$ signal into ISHE (red curve) and galvanomagnetic effects (black curve) based on this ratio is shown in Fig. \ref{fig5}(c).

\begin{figure}[t]
\centerline{\includegraphics[width=14cm, angle=0, bb= 0 0 460 260]{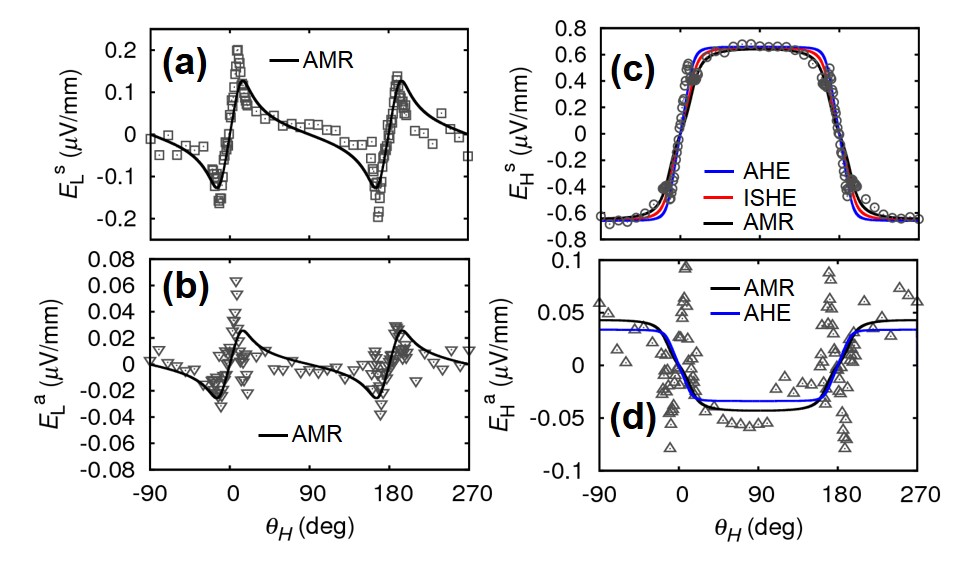}}
\caption{\label{fig6}
Measured angular dependences of the four components (a) $E_{\text{L}}^{\text{s}}$, (b) $E_{\text{L}}^{\text{a}}$, (c) $E_{\text{H}}^{\text{s}}$, and (d) $E_{\text{H}}^{\text{a}}$ for a La$_{0.67}$Sr$_{0.33}$MnO$_{3}$$\mid$ Pt bilayer wired as in Fig. \ref{fig4}. The black, red and blue curves are the calculated curves for AMR, ISHE and AHE, respectively (not to scale). 
}
\end{figure}

To corroborate our findings in Ni$_{18}$Fe$_{19}$$\mid$Pt samples, we show results on a La$_{0.67}$Sr$_{0.33}$MnO$_{3}$$\mid$ Pt bilayer sample in Fig. \ref{fig6}. The La$_{0.67}$Sr$_{0.33}$MnO$_{3}$ layer (20 nm) has been prepared by pulsed laser deposition on SrTiO$_3$ substrate and covered with Pt (10 nm) by sputtering. The wiring of the sample in this case was identical to that in Fig. \ref{fig4}. The $Q$-value during measurement was 7200, and the microwave power $P_{\text{MW}}=20$ mW (rf field $\mu_0 h=0.05$ mT). The experimental data in Fig. \ref{fig6} (grey points) are plotted along with the calculated curves for AMR, AHE and ISHE, with $\mu_0 M_{\text{S}}=0.570$ T and $\omega/\gamma=0.344$ T determined from the angular dependence of $H_{\text{R}}$, and $\alpha = 0.02$. There is a good agreement between the calculation and the experiment in the $E_{\text{L}}^{\text{s}}$ and the $E_{\text{L}}^{\text{a}}$ components [Figs. \ref{fig6} (a), (b)]. Some discrepancies can be seen near sign reversal in $E_{\text{H}}^{\text{s}}$ and $E_{\text{H}}^{\text{a}}$ [Figs. \ref{fig6} (c), (d)]. Significantly, the $E_{\text{H}}^{\text{s}}$ component is much larger than the remaining three components. We obtain $E_{\text{H}}^{\text{a}}/E_{\text{L}}^{\text{a}}=1.45$ and $E_{\text{H}}^{\text{s}}/E_{\text{L}}^{\text{s}}=3.43$, showing that the electromotive force component including ISHE is more than three times larger than the other three components.

\section{Conclusions}

We have calculated and measured the angular dependences of electromotive forces induced in Ni$_{81}$Fe$_{19}$ and Ni$_{18}$Fe$_{19}$$\mid$Pt thin films at ferromagnetic resonance in a microwave cavity. The predicted contributions from the inverse spin Hall effect, anomalous Hall effect and anisotropic magnetoresistance exhibit distinct signal symmetry as well as angular dependence when changing the direction of the external field with respect to the film plane. 
The results of the calculation indicate that a complete separation of the inverse spin Hall effect from galvanomagnetic effects as well as a determination of the phase shift between the rf electric and magnetic fields in the cavity is in principle possible by measuring the angular dependence in both longitudinal and Hall directions. 
According to spin pumping measurements on the above systems, only in Ni$_{81}$Fe$_{19}$$\mid$Pt  is the component including the inverse spin Hall effect more than five times larger than components consising only of galvanomagnetic effects. This indicates that the inverse spin Hall effect is dominant in the observed symmetric signal. This observation is supported by results on La$_{0.67}$Sr$_{0.33}$MnO$_{3}$$\mid$ Pt bilayers. However, the angular dependences of the electromotive force components in the longitudinal direction include contributions not expected in a TE$_{011}$ cavity, pointing to the possibility that the distribution of microwave electromagnetic fields in the cavity is altered by the presence of the sample and its wiring.

\begin{acknowledgments}
This work was supported by CREST
``Creation of Nanosystems with Novel Functions through Process Integration” from JST, Japan, 
and Grants-in-Aid for Scientific Research on Innovative Areas 
Nano Spin Conversion Science (No. 26103005), 
Challenging Exploratory Research (No. 26610091) and 
Scientific Research (A) (No. 24244051) from MEXT, Japan.
\end{acknowledgments}



\begin{thebibliography}{00}
\addcontentsline{toc}{chapter}{References}

\bibitem{mizukami} S. Mizukami, Y. Ando, and T. Miyazaki, Phys. Rev. B \textbf{66}, 104413 (2002).
\bibitem{tserkovnyak} Y. Tserkovnyak, A. Brataas, and G. E. W. Bauer, Phys. Rev. Lett. \textbf{88}, 117601 (2002).
\bibitem{saitoh} E. Saitoh, M. Ueda, H. Miyajima, and G. Tatara, Appl. Phys. Lett. \textbf{88}, 182509 (2006).
\bibitem{azevedo-ishe} A. Azevedo, L. H. Vilela Leao, R. L. Rodriguez-Suarez, A. B. Oliveira, and S. M. Rezende, J. Appl. Phys. \textbf{97}, 10C715 (2005).
\bibitem{costache} M. V. Costache, M. Sladkov, S. M. Watts, C. H. van der Wal, and B. J. van Wees, Phys. Rev. Lett. \textbf{97}, 216603 (2006).
\bibitem{kajiwara} Y. Kajiwara, K. Harii, S. Takahashi, J. Ohe, K. Uchida, M. Mizuguchi, H. Umezawa, H. Kawai, K. Ando, K. Takanashi, S. Maekawa, and E. Saitoh, Nature \textbf{464}, 262 (2010).
\bibitem{ando-prb} K. Ando, Y. Kajiwara, S. Takahashi, S. Maekawa, K. Takemoto, M. Takatsu, and E. Saitoh, Phys. Rev. B \textbf{78}, 014413 (2008).
\bibitem{ando} K. Ando, S. Takahashi, J. Ieda, Y. Kajiwara, H. Nakayama, T. Yoshino, K. Harii, Y. Fujikawa, M. Matsuo, S. Maekawa, and E. Saitoh, J. Appl. Phys. \textbf{109}, 103913 (2011).
\bibitem{mcguire} T. R. McGuire and R. I. Potter, IEEE Trans. Magn. \textbf{11}, 1018 (1975).
\bibitem{nagaosa} N. Nagaosa, J. Sinova, S. Onoda, A. H. MacDonald, and N. P. Ong, Rev. Mod. Phys. \textbf{82}, 1539 (2010).
\bibitem{jan} J.-P. Jan, Solid State Phys. \textbf{5}, 1 (1957).

\bibitem{LC} L. Chen, F. Matsukura, and H. Ohno, Nature Comm. \textbf{4}, 2055 (2013).

\bibitem{harder} M. Harder, Z. X. Cao, Y. S. Gui, X. L. Fan, and C.-M. Hu, Phys. Rev. B, \textbf{84}, 054423 (2011). 
\bibitem{azevedo} A. Azevedo, L. H. Vilela-Leao, R. L. Rodriguez-Suarez, A. F. Lacerda Santos, and S. M. Rezende, Phys. Rev. B \textbf{83}, 144402 (2011).
\bibitem{juretschke-1960} H. J. Juretschke, J. Appl. Phys. \textbf{31}, 1401 (1960).
\bibitem{juretschke-1962} W.G. Egan and H. J. Juretschke, J. Appl. Phys., \textbf{34}, 1477 (1963).
\bibitem{bai} L. Bai, P. Hyde, Y. S. Gui, C.-M. Hu, V. Vlaminck, J. E. Pearson, S. D. Bader, and A. Hoffmann, Phys. Rev. Lett. \textbf{111}, 217602 (2013).
\bibitem{LC-apex} L. Chen, S. Ikeda, F. Matsukura, and H. Ohno, Appl. Phys. Exp. \textbf{7}, 013002 (2014).

\bibitem{obstbaum} M. Obstbaum, M. H{\"a}rtinger, H. G. Bauer, T. Meier, F. Swientek, C. H. Back, and G. Woltersdorf, Phys. Rev. B \textbf{89}, 060407(R) (2014).
\bibitem{weetee} W. T. Soh, B. Peng, and C. K. Ong, J. Phys. D: Appl. Phys. \textbf{47}, 285001 (2014).
\bibitem{haidar} S. M. Haidar, R. Iguchi, A. Yagmur, J. Lustikova, Y. Shiomi, and E. Saitoh, J. Appl. Phys. \textbf{117}, 183906 (2015).
\bibitem{poole} C. Poole, \textit{Electron Spin Resonance: A Comprehensive Treatise on Experimental Techniques}, Willey (New York, 1983).
\bibitem{mecking} N. Mecking, Y. S. Gui, and C.-M. Hu, Phys. Rev. B \textbf{76}, 224430 (2007).

\bibitem{ciccarelli} Chiara Ciccarelli, Kjetil M. D. Hals, Andrew Irvine, Vit Novak, Yaroslav Tserkovnyak, Hidekazu Kurebayashi, Arne Brataas, and Andrew Ferguson, Nature Nanotechnology \textbf{10}, 50 (2014).
\bibitem{azevedo2015} A. Azevedo, R. O. Cunha, F. Estrada, O. Alves Santos, J. B. S. Mendes, L. H. Vilela-Leao, R. L. Rodriguez-Suarez, and S. M. Rezende, Phys. Rev. B \textbf{92}, 024402 (2015)


\end{thebibliography}
\end{document}